# A COMPREHENSIVE STUDY OF NATURAL GAMMA RADIOACTIVITY LEVELS AND ASSOCIATED DOSE RATES FROM SURFACE SOILS IN CYPRUS

M. Tzortzis, E. Svoukis and H. Tsertos[*]

*Department of Physics, University of Cyprus, P.O. Box 20537, 1678 Nicosia, Cyprus.*

**(Revised version: 2/03/2004)**

## Abstract

A survey was carried out to determine activity concentration levels and associated dose rates from the naturally occurring radionuclides $^{232}$Th, $^{238}$U, and $^{40}$K, in the various geological formations of Cyprus, by means of high-resolution γ–ray spectrometry. A total of 115 representative soil samples were collected from all over the bedrock surface of the island, based on the different lithological units of the study area, sieved through a fine mesh, sealed in 1000-*mL* plastic Marinelli beakers, and measured in the laboratory with respect to their gamma radioactivity for a counting time of 18 *hours* each. From the measured spectra, activity concentrations were determined for $^{232}$Th (range from $1.0 \times 10^{-2}$ to 39.8 *Bq kg$^{-1}$*), $^{238}$U (from $1.0 \times 10^{-2}$ to 39.3 *Bq kg$^{-1}$*) and $^{40}$K (from $4.0 \times 10^{-2}$ to 565.8 *Bq kg$^{-1}$*). Gamma absorbed dose rates in air outdoors were calculated to be in the range of $1.1 \times 10^{-2}$ – 51.3 *nGy h$^{-1}$*, depending on the geological features, with an overall mean value of 8.7 *nGy h$^{-1}$* and a standard deviation of 8.4 *nGy h$^{-1}$*. This value is by a factor of about 7 below the corresponding population-weighted world-averaged value of 60 *nGy h$^{-1}$* and one of the lowest that has been reported from similar investigations worldwide. Assuming a [*] **Corresponding author**. *E-mail address: tsertos@ucy.ac.cy, Fax: +357-22892821*.

20% occupancy factor, the corresponding effective dose rates outdoors equivalent to the population were calculated to be between $1.3\times10^{-2}$ and 62.9 $\mu Sv\ y^{-1}$, with an arithmetic mean value of 10.7 $\mu Sv\ y^{-1}$ and a standard deviation of 10.3 $\mu Sv\ y^{-1}$.

# INTRODUCTION

The great interest expressed worldwide for the study of naturally occurring radiation and environmental radioactivity has led to the performance of extensive surveys in many countries[1]. Such investigations can be useful for both the assessment of public dose rates and the performance of epidemiological studies, as well as to keep reference-data records, to ascertain possible changes in the environmental radioactivity due to nuclear, industrial, and other human activities.

Natural environmental radioactivity arises mainly from primordial radionuclides, such as $^{40}K$, and the radionuclides from the $^{232}Th$ and $^{238}U$ series and their decay products, which occur at trace levels in all ground formations. Primordial radionuclides are formed by the process of nucleosynthesis in stars and are characterised by half-lives comparable to the age of the earth.

Gamma radiation emitted from those naturally occurring radioisotopes, also called terrestrial background radiation, represents the main external source of irradiation of the human body. Natural environmental radioactivity and the associated external exposure due to gamma radiation depend primarily on the geological and geographical conditions, and appear at different levels in the soils of each region in the world[1]. The specific levels of terrestrial environmental radiation are related to the geological composition of each lithologically separated area, and to the content in thorium (Th), uranium (U) and potassium (K) of the rock from which the soils originate in each area. In terms of natural radioactivity, it is well known, for instance, that igneous rocks of granitic composition are strongly enriched in Th and U (on an average 15 *ppm*[†] of Th and 5 *ppm* of U), compared to rocks of basaltic or ultramafic composition (< 1 *ppm* of U)[2,3]. For that reason, higher radiation levels are associated with igneous rocks and lower levels with sedimentary rocks. There are exceptions, however, as some shales and phosphate rocks have relatively high content of those radionuclides[1].

The island of Cyprus is located in the eastern basin of the Mediterranean Sea and extends to an area of about 9,300 $km^2$. The characteristic geological formations can be classified into two main categories: those that belong to an ophiolite complex and those of sedimentary origin. The Cyprus ophiolite is one of the best-preserved and most intensively studied ophiolite complexes in the world, and is known as the Troodos Massif or Troodos Ophiolitic Complex[4,5]. It consists of basic and ultrabasic pillow lavas, fringed by andesitic-sheeted dykes, while the central part of the ophiolite is composed of basic and ultrabasic plutonic rocks (gabbros, peridotites, dunites and serpentinised harzburgites). The weathering of the rocks of sedimentary origin (mainly chalks, marls, sandstones, gypsum, etc.) in the foothills that fringe Troodos mountains, gave rise to a diversity of alkaline, calcium-rich soils[5]. On the other hand, rocks of ophiolitic origin are mostly iron- and magnesium-saturated or oversaturated, and this allows their characterization as basic and ultrabasic. None of these rock types belongs to the category of silica-oversaturated, which usually is associated with high Th and U elemental concentrations[2].

---

[†] Parts per million

With the objective to obtain the first systematic data on environmental radiation and radioactivity for Cyprus, a pilot project was commenced in 2001. The first results of a feasibility study were based on 28 samples collected from the main geological rock types appearing in the island[6]. Other parts of the project aimed at measuring the concentration of radon in houses and public buildings[7], as well as at detecting α−emitting radioisotopes by means of radio-analytical techniques and high-resolution α-spectrometry[8].

A follow-up, extensive and completed study of terrestrial gamma radioactivity was performed after the systematic collection of 115 soil samples over all the island bedrock surface based on the different lithological units of the study area. Such a number of collected samples in combination with the relatively large area covered by this study (about 6,000 $km^2$) can be considered as a representative inventory of the various outcropping geological formations. This investigation becomes particularly important and interesting to radiometric studies, considering that it provides information on the geomorphological composition and associated environmental radioactivity of such an area with a large variety of ophiolitic and sedimentary rock types cropping out over a relatively narrow area. In this paper, the results from these extensive investigations regarding activity concentrations and associated gamma dose rates due to naturally occurring radionuclides in a wide variety of surface soils are presented. The experiments have been carried out in the Nuclear Physics Laboratory of the Department of Physics, University of Cyprus, using a high-resolution γ−ray spectrometry system.

## MATERIALS AND METHODS

**Sample collection and preparation**

A total of 115 surface soil samples has been collected throughout the geological surface formations in the accessible area of the island‡ (Figure 1). The official geological map of Cyprus[9] was used to segregate the area studied into seven different geological regions, as indicated in the simplified map shown in Figure 1. Ten to 28 soil samples were collected from different locations falling within the boundaries of each of the seven geological regions considered. More than half (59) of the total of samples was collected from the Troodos Ophiolitic Complex.

Soil samples were collected at 115 different sites with the only constraint that no sampling site should be taken close to a field boundary, a road, a tree, a building, or other obstruction. For each soil sample collected, an area of about 0.5 $m$ × 0.5 $m$ was marked and carefully cleared of debris to a few centimetres depth. Surface soils were then taken from different places randomly within the marked and cleared area, and mixed together thoroughly, in order to obtain a representative sample of that area. Each soil sample was labelled according to the geographical coordinates of the sampling area, and those coordinates were later used to indicate the position of the sampling area on the simplified map by open circle points (see Figure 1).

Collected soil samples were air-dried, sieved through a fine mesh (~0.5 $mm$), hermetically sealed in standard 1000-$mL$ plastic Marinelli beakers, dry-weighed and stored for about four weeks prior to counting. This allowed to obtain on the one hand samples in which the rather large soil mass used (being typically between 1.0 and 1.4

---

‡ Samples from the northern part of Cyprus could not be collected, because of the political division of the island since 1974 (see Figure 1).

*kg*) was greatly homogenised within the beaker volume and to ensure on the other hand that equilibrium between $^{226}$Ra and $^{222}$Rn and their decay products was established. Concerning the last point one should notice that equilibrium is common in rocks older than $10^6$ *years*, and the $^{232}$Th series may be considered in equilibrium in most geological environments[10].

**Sample counting**

Each prepared soil container was placed on a shielded, high-purity germanium (HPGe) detector that is characterised by 33% efficiency, relative to a 3″×3″ NaI(Tl) scintillator, and by 1.8 *keV* energy resolution (FWHM) at the 1.33 *MeV* reference transition of $^{60}$Co. The samples were measured for a counting time of 18 *hours*, to obtain a statistically small error (3-5%) for the γ–ray peaks of interest. Further details of the high-resolution spectrometry system, as well as of the data analysis technique are presented elsewhere[6,11].

Following the spectrum analysis, count rates for each detected photopeak and activity concentration in units of *Bq kg$^{-1}$* for each of the detected nuclides are calculated[6]. The total uncertainty of the radioactivity measurements, which is also applicable to the calculated gamma dose and effective dose rates, was typically in the range 3−10%. It has been calculated by taking into consideration the counting statistical error (~3%) and other weighted systematic errors that mainly include the uncertainty in the efficiency calibration (0.5–8%)[11]. Depending on the peak background of the measured spectra, the Minimum Detectable Activity (MDA) was calculated to be $1.0\times10^{-2}$ *Bq kg$^{-1}$* for both $^{232}$Th and $^{238}$U, and $4.0\times10^{-2}$ *Bq kg$^{-1}$* for $^{40}$K, for the counting time of 18 *hours.*

# RESULTS AND DISCUSSION

For the purpose of quality assurance, a control sample from a reference soil material (IAEA-326) was prepared in an identical 1000-mL Marinelli beaker (dry soil mass of 1.254 kg), and treated with respect to the measurement and analysis procedure as an unknown sample. The results obtained and the certified values[12] are summarised in Table 1. As can be seen, the certified mean values of the $^{232}$Th, $^{238}$U, and $^{40}$K activity concentrations are reproduced by the measurements with a relative deviation of ~6%, ~8%, and ~1%, respectively, demonstrating a rather good performance of the measurement and analysis technique utilised. Also, the measured activity concentration values from the soil samples presented have been checked regarding their reproducibility and consistency by repeating the measurement and analysis procedure for three different samples from each of the 7 regions considered. The samples of the repetitive measurements were selected randomly among those initially collected and stored, after a time of 9 to 12 months from their first measurement. In all cases examined, the measured activities of the first measurement could be reproduced by the repeated measurement within 5 to 15%.

The calculated activity concentrations of $^{232}$Th, $^{238}$U and $^{40}$K for the 115 samples collected from all over the island bedrock surface are plotted in Figure 2. As can be seen, higher values of activity concentrations are associated with formations of sedimentary origin (regions R1, R2, R3 and R5) and lower values are associated with formations belonging to the Troodos ophiolitic complex (regions R4, R6 and R7). In particular, region R4, which covers the central part of the Troodos massif that contains basic and ultrabasic plutonic rocks (see Figure 1), presents the lowest activity concentrations of $^{232}$Th, $^{238}$U, and $^{40}$K.

Activity concentrations of $^{232}$Th ranged from $1.0 \times 10^{-2}$ to 39.8 *Bq kg$^{-1}$*, from $1.0 \times 10^{-2}$ to 39.3 *Bq kg$^{-1}$* for $^{238}$U, and from $4.0 \times 10^{-2}$ to 565.8 *Bq kg$^{-1}$* for $^{40}$K. It should be noted that the lower values given correspond to the detection limit (MDA) derived for each of the three radionuclides, since in some samples (mainly from region R4) the corresponding net activity concentration (after the subtraction of the laboratory ambient background) exhibited nearly zero or even negative values. In general, activity concentrations of $^{232}$Th and $^{238}$U are rather low in all the samples measured; this is particularly observable in samples from ophiolitic formations, which contain the lowest concentrations of those radionuclides.

In Table 2, averaged concentrations for each one of the seven geological regions studied are presented. The highest mean activity concentrations of $^{232}$Th, $^{238}$U and $^{40}$K were measured in soil samples derived from sedimentary formations, while the mean activity concentrations associated with soils that stem from the Troodos ophiolitic complex are significantly lower. To be more precise, the average activity concentration of $^{232}$Th, $^{238}$U, and $^{40}$K in samples of sedimentary origin ranged from 5.7 to 16.7 *Bq kg$^{-1}$* with a mean of 9.3 ± 8.1 *Bq kg$^{-1}$*, from 9.6 to 21.9 *Bq kg$^{-1}$* with a mean of 13.2 ± 8.8 *Bq kg$^{-1}$*, and from 116.8 to 190.2 *Bq kg$^{-1}$* with a mean of 146.9 ± 97.6 *Bq kg$^{-1}$*, respectively. The corresponding activity concentration values in soils of ophiolitic origin ranged from 0.3 to 1.8 *Bq kg$^{-1}$* with a mean of 1.0 ± 1.4 *Bq kg$^{-1}$*, from 0.5 to 2.5 *Bq kg$^{-1}$* with a mean of 1.3 ± 1.3 *Bq kg$^{-1}$*, and from 27.8 to 158.2 *Bq kg$^{-1}$* with a mean of 64.4 ± 72.0 *Bq kg$^{-1}$*, respectively.

Arithmetic mean values calculated from all samples of both sedimentary and ophiolitic origin are: 5.0 ± 7.1, 7.1 ± 8.6, and 104.6 ± 94.6 *Bq kg$^{-1}$*, for $^{232}$Th, $^{238}$U and $^{40}$K activity concentrations, respectively, while median values derived from all data available worldwide[1] are: 30, 35, and 400 *Bq kg$^{-1}$*, respectively. This reveals that the mean activity concentration levels measured in Cyprus from naturally occurring radioisotopes are by a factor of four to six lower than the corresponding values obtained worldwide. Region 5, which is mainly composed of calcarenites and sandstones, shows the highest activity concentration in $^{232}$Th and $^{40}$K radionuclides with mean values of 16.7 ± 9.4 and 190.2 ± 94.7 *Bq kg$^{-1}$*, respectively, while Region 2, with soils developed from chalks, marls and gypsum, exhibits the highest activity concentration in $^{238}$U with a mean of 21.9 ± 10.2 *Bq kg$^{-1}$*. The highest mean activity concentrations of $^{232}$Th, $^{238}$U and $^{40}$K are still well below the corresponding worldwide mean activity concentration values reported in the UNSCEAR 2000 Report [1].

The gamma absorbed dose rates in air outdoors were calculated from the concentrations of each of the nuclides of $^{232}$Th and $^{238}$U series, and of $^{40}$K, in a similar manner than that described in details by Tzortzis et al. (2003)[6]. The total dose rates in air outdoors calculated for samples of sedimentary origin ranged from 1.1 to 51.3 *nGy h$^{-1}$* with a mean of 14.3 ± 8.3 *nGy h$^{-1}$* and from $1.1 \times 10^{-2}$ to 16.9 *nGy h$^{-1}$* with a mean of 3.3 ± 3.5 *nGy h$^{-1}$* for samples of ophiolitic origin. The arithmetic mean value of the total absorbed dose rate over all samples amounts to 8.7 *nGy h$^{-1}$* with a standard deviation of 8.4 *nGy h$^{-1}$*. The results of the calculated total gamma absorbed dose in air outdoors due to the $^{232}$Th, $^{238}$U, and $^{40}$K radionuclides are summarised in Figure 3, while in Figure 4 the frequency distribution of the total absorbed dose rates for all the measured 115 soil samples is plotted.

Figure 5 shows the calculated relative contribution to the total absorbed dose of each of the three radionuclides ($^{232}$Th, $^{238}$U, and $^{40}$K) for the seven main regions considered in this study (see Figure 1). As illustrated in Figure 5, the contribution to total dose rates due to $^{40}$K, $^{238}$U and $^{232}$Th series for soil samples of sedimentary origin (regions R1, R2, R3 and R5) is more or less equally shared between those three radionuclides (~39%, ~36% and ~25%, respectively). On the other hand, in soils of ophiolitic origin (regions R4, R6, and R7) $^{40}$K contributes about 75% on an average of the total absorbed dose, which is significantly higher than those due to $^{238}$U and $^{232}$Th series (~15% and ~10%, respectively).

Finally, the effective dose rates outdoors estimated for soil samples of sedimentary and ophiolitic origin ranged from 1.3 to 62.9 *µSv y$^{-1}$* with a mean of 17.6 ± 10.2 *µSv y$^{-1}$* and from 1.3×10$^{-2}$ to 20.7 *µSv y$^{-1}$* with a mean of 4.1 ± 4.3 *µSv y$^{-1}$*, respectively. As a consequence, the Cypriot population living outdoors is subjected to a mean effective dose rate of 10.7 ± 10.3 *µSv y$^{-1}$* due to naturally occurring gamma radiation.

The results presented in this paper are in general agreement with the first measurements of activity concentrations and calculated associated dose rates of naturally occurring radionuclides in 28 samples from characteristic geological rock types of Cyprus[6]. In that study, soil samples of sedimentary composition appeared again with higher activity concentration levels of the radioisotopes of interest than those exhibited by soils of ophiolitic origin as following: 16.7 ± 3.7 versus 2.1 ± 0.2 *Bq kg$^{-1}$* for $^{232}$Th, 21.0 ± 5.5 versus 6.5 ± 4.4 *Bq kg$^{-1}$* for $^{238}$U, and 197.5 ± 35.2 versus 147.3 ± 79.3 *Bq kg$^{-1}$* for $^{40}$K, for samples of sedimentary and ophiolitic origin, respectively. Arithmetic mean values of the measured activity concentrations over all samples were found to be 11.0 ± 13.8, 15.4 ± 20.8, and 177.8 ± 196.4 *Bq kg$^{-1}$*, for $^{232}$Th, $^{238}$U and $^{40}$K, respectively. The total absorbed dose rates were equal to 19.7 ± 3.8 and 7.4 ± 4.1 *nGy h$^{-1}$*, for samples of sedimentary and ophiolitic origin, respectively, with an overall mean value of 14.7 ± 7.3 *nGy h$^{-1}$*. One should notice in this context that in the present study more than half (∼51%) of the samples collected are of ophiolitic origin, which revealed very low activity concentration levels compared to those of sedimentary origin. The corresponding percentage contribution of samples of ophiolitic origin in the first study[6] was ∼39%, and this might explain why the results of the activity concentration and dose rates are slightly higher compared to the corresponding values of the present study.

In Table 3, a summary of recent results on natural gamma radioactivity levels derived from similar investigations conducted in regions close and around the Mediterranean Sea is presented. As can be seen, the corresponding activity concentration levels of the $^{232}$Th, $^{238}$U, and $^{40}$K radioisotopes obtained from this study fall within the lowest range of most reported values from other worldwide and neighbouring areas. In other words, since natural radioactivity is mainly composed of these three radioisotopes, the island of Cyprus can be considered as one of the world areas that exhibit very low levels of natural radioactivity.

**CONCLUSION**

High-resolution γ–ray spectrometry was exploited to determine activity concentration and the associated dose rates due to naturally occurring $^{232}$Th, $^{238}$U, and $^{40}$K radioisotopes in 115 surface soil samples collected from all over the island of Cyprus. Soils originated from the Troodos ophiolitic complex appear generally to present lower naturally occurring radionuclide concentrations, compared to those of sedimentary origin. In the present study, arithmetic mean activity concentrations (mean ± S.D.) of 5.0 ± 7.1, 7.1 ± 8.6 and 104.6 ± 94.6 *Bq kg$^{-1}$*, for $^{232}$Th, $^{238}$U and $^{40}$K radionuclides, respectively, are derived from all the soil samples studied. These values fall within the lowest range of those measured at worldwide scale by other authors and, more specifically, are by a factor of four to six lower than the world average values[1] of 30, 35, and 400 *Bq kg$^{-1}$*, for $^{232}$Th, $^{238}$U, and $^{40}$K, respectively.

Gamma absorbed dose rates in air outdoors were calculated to be in the range of $1.1 \times 10^{-2}$ – 51.3 *nGy h$^{-1}$*, with an overall mean value of 8.7 *nGy h$^{-1}$* and a standard deviation of 8.4 *nGy h$^{-1}$*. This value is by a factor of about 7 below the corresponding population-weighted world-averaged value of 60 *nGy h$^{-1}$* and one of the lowest that has been reported from similar investigations worldwide[1]. As a direct implication of these results, inhabitants of the island are subjected to an external gamma radiation exposure (effective dose outdoors) that range from $1.3 \times 10^{-2}$ to 62.9 *μSv y$^{-1}$* with an arithmetic mean value of 10.7 ± 10.3 *μSv y$^{-1}$*, which is negligibly small compared to the worldwide average exposure of ~0.7 *mSv y$^{-1}$* (~30% × 2.4 *mSv y$^{-1}$*) due to gamma radiation from naturally occurring radioisotopes[1].

The results obtained from the present wide-range investigations are in general agreement with the first studies on activity and elemental concentrations of naturally occurring radionuclides and on their associated dose rates in various rock types[6]. They further confirm that the activity concentration and associated dose rates due to naturally occurring gamma radiation in Cyprus are significantly lower (by a factor of five on an average) than the corresponding radioactivity levels reported from other areas worldwide.

## Acknowledgements

This work is financially supported by the Cyprus Research Promotion Foundation (Grant No. 45/2001), and the University of Cyprus.

**TABLE CAPTIONS**

**Table 1.** Quality control measurements using the reference soil material IAEA-326 that has been treated as an "unknown" sample.

**Table 2.** Average activity concentration and standard deviations of naturally occurring $^{232}$Th, $^{238}$U and $^{40}$K radionuclides in soil samples from the main geological regions studied (see Figure 1).

**Table 3.** Summary of activity concentration ($Bq\ kg^{-1}$) of naturally occurring $^{232}$Th, $^{238}$U, and $^{40}$K radioisotopes in soil samples from work conducted in regions around the Mediterranean Sea.

**FIGURE CAPTIONS**

**Figure 1**. Simplified geological map of Cyprus, indicating the seven geological regions examined in this survey; the various rock formations are described in the legend. Open circle points indicate the locations from where soil samples were collected in each geological region.

**Figure 2.** $^{232}$Th, $^{238}$U, and $^{40}$K activity concentrations for all the soil samples studied. The vertical dashed lines correspond to the seven geological regions (R1-R7) from which the measured samples have been collected (Figure 1).

**Figure 3.** Total absorbed dose rate in air outdoors due to gamma radiation for all the measured soil samples. The horizontal bold line represents the overall arithmetic mean value of 8.7 *nGy h$^{-1}$*.

**Figure 4.** Frequency distribution of the total absorbed dose rate in air outdoors due to gamma radiation for all the measured soil samples.

**Figure 5.** $^{232}$Th, $^{238}$U, and $^{40}$K percentage contribution to the total absorbed dose rate in air outdoors for the seven main geological regions studied (see Figure 1).

**Table 1.**

### Reference soil material IAEA-326

(Dry soil mass used: 1.254 *kg*)

| Radionuclide | Certified activities *(Bojanowski et al., 2001)*[12] | | Measurement results[**] |
|---|---|---|---|
| | **Mean value** $(Bq\ kg^{-1})$ | **95% Confidence interval** $(Bq\ kg^{-1})$ | **Activity ± S.D.** $(Bq\ kg^{-1})$ |
| $^{232}$Th | 39.4 | 37.6 – 41.2 | 41.5 ± 0.6 |
| $^{238}$U | 29.4 | 28.1 – 30.7 | 31.8 ± 0.5 |
| $^{40}$K | 580 | 571 – 589 | 575 ± 6 |

[**] Measurements were taken at three times on the reference soil.

**Table 2.**

| | Region | Number of samples | Average activity concentration ($Bq\ kg^{-1}$) | | | | | |
|---|---|---|---|---|---|---|---|---|
| | | | $^{232}Th$ | | $^{238}U$ | | $^{40}K$ | |
| | | | Mean | S. D. | Mean | S. D. | Mean | S. D. |
| **Sedimentary Rocks** | R1 | 23 | 5.7 | 8.1 | 9.6 | 7.2 | 141.1 | 126.1 |
| | R2 | 14 | 11.1 | 6.0 | 21.9 | 10.2 | 150.4 | 65.8 |
| | R3 | 10 | 8.5 | 3.5 | 10.1 | 4.3 | 116.8 | 45.9 |
| | R5 | 9 | 16.7 | 9.4 | 12.0 | 4.2 | 190.2 | 94.7 |
| | Total | 56 | 9.3 | 8.1 | 13.2 | 8.8 | 146.9 | 97.6 |
| **Ophiolitic Rocks** | R4 | 28 | 0.3 | 0.7 | 0.5 | 0.5 | 27.8 | 27.3 |
| | R6 | 21 | 1.5 | 1.6 | 1.8 | 1.1 | 68.6 | 45.6 |
| | R7 | 10 | 1.8 | 1.4 | 2.5 | 1.7 | 158.2 | 112.2 |
| | Total | 59 | 1.0 | 1.4 | 1.3 | 1.3 | 64.4 | 72.0 |
| | Grand Total | 115 | 5.0 | 7.1 | 7.1 | 8.6 | 104.6 | 94.6 |

**Table 3.**

| Region | $^{232}$Th ($Bq\ kg^{-1}$) | $^{238}$U ($Bq\ kg^{-1}$) | $^{40}$K ($Bq\ kg^{-1}$) | Reference |
|---|---|---|---|---|
| Amman, Jordan | 28.8 | 56.4 | 501.3 | Ahmad et al. (1997) [13] |
| Karak, Jordan | 27.2 | 228.9 | 410.2 | Ahmad et al. (1997) [13] |
| Instanbul, Turkey | 37 | 21 | 342 | Karahan and Bayulken (2000) [14] |
| Costal area, Aegean sea, Greece | 71 ± 25 | 93 ± 47 | 877 ± 352 | Florou and Kritidis (1992) [15] |
| Russaifa, Jordan | 8.7–27.1 | 48.3–523.2 | 44–307 | Al-Jundi (2002) [16] |
| Italy | 73–87 | 57–71 | 580–760 | Bellia et al. (1997) [17] |
| Spain | 13.2–84.4 | 20.3–71.1 | 289–703 | Martinez-Aguirre and Garcia-Leon (1997) [18] |
| Canary islands | 54 | 44 | 665 | Fernandez et al. (1992) [19] |
| Spain | 7–204 | 13–165 | 48–1570 | Baeza et al. (1992) [20] |
| Cyprus average | 5.0 ± 7.1 | 7.1 ± 8.6 | 104.6 ± 94.6 | Present study |
| Worldwide average | 30 | 35 | 400 | UNSCEAR report (2000) [1] |

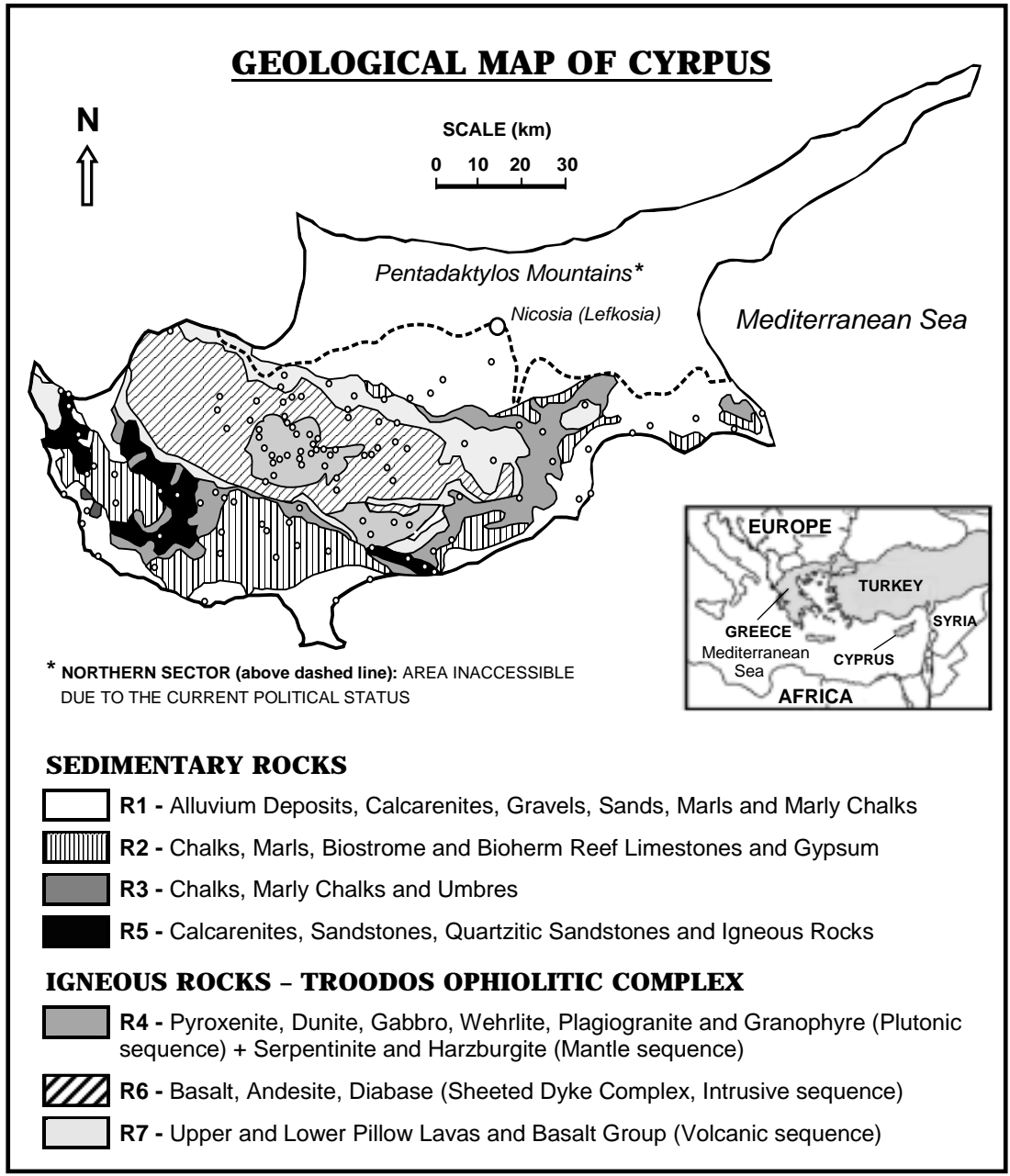

**Figure 1.**

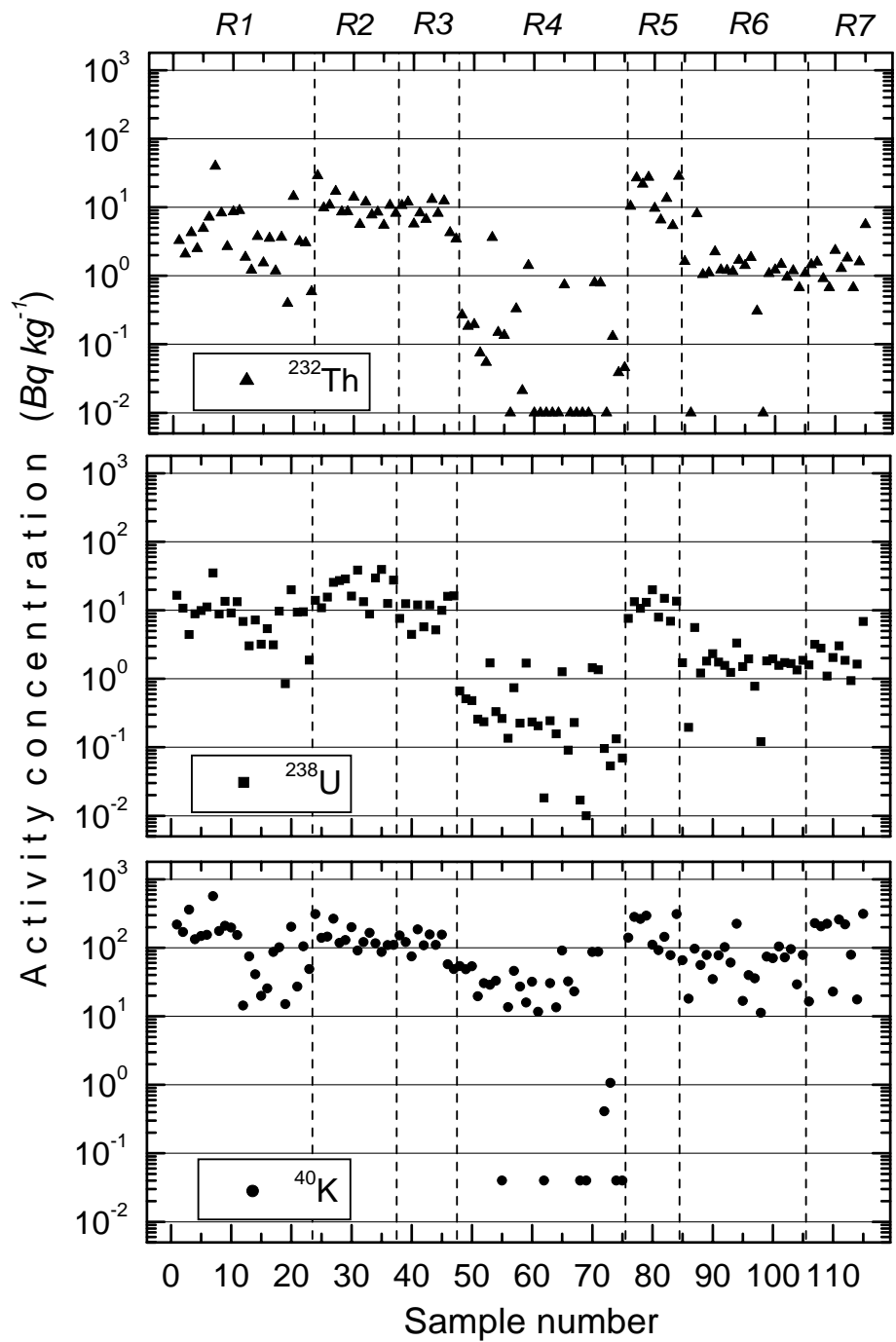

**Figure 2.**

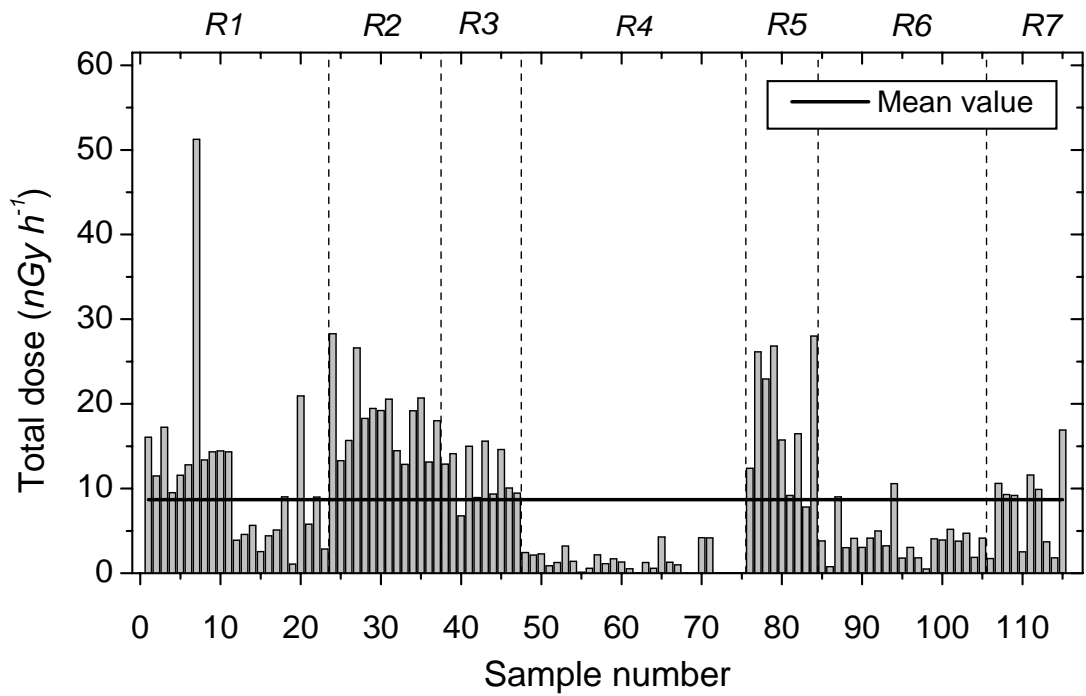

**Figure 3.**

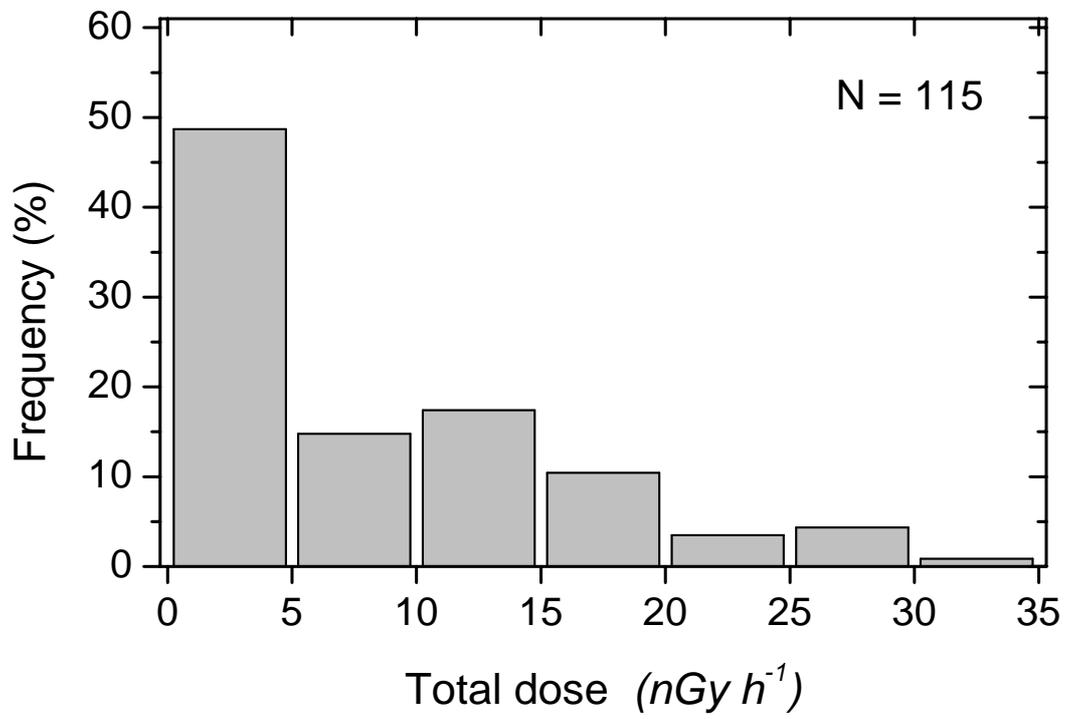

**Figure 4.**

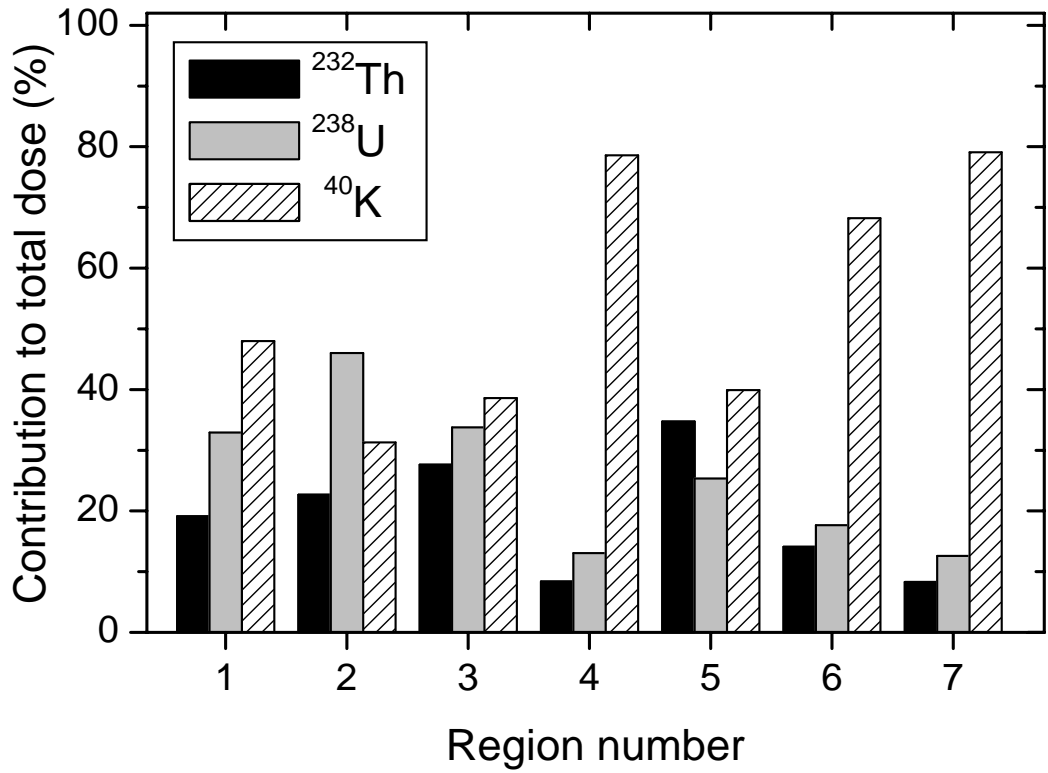

**Figure 5.**